\documentstyle[12pt]{article}
\def\be{\begin{equation}}
\def\ee{\end{equation}}
\def\bea{\begin{eqnarray}}
\def\eea{\end{eqnarray}}

\begin{document}
\begin{center}
{\Large{\bf Attached Open Strings on a D$p$-brane
in the Backgrounds of the pp-wave and Linear Dilation}}

\vskip .5cm
{\large Davoud Kamani}
\vskip .1cm
{\it Physics Department,
Amirkabir University of Technology (Tehran Polytechnic)\\}
P.O.Box: {\it 15875-4413, Tehran, Iran}\\
e-mail: {\it kamani@aut.ac.ir}\\
\end{center}

\begin{abstract}

Open strings on a D$p$-brane in the
pp-wave spacetime, accompanied by a linear dilaton background,
will be studied. Various properties of this 
system such as solvability of equations of 
motion and quantization will be investigated.

\end{abstract}
\vskip .5cm



\newpage
\section{Introduction}

Various backgrounds have been applied for revealing the 
hidden properties of the string theory.
Only some of them admit solvability of the string theory. One of
these backgrounds is the pp-wave metric which is 
supported by a constant 5-form flux \cite{1}, and can be derived from
the $AdS_5 \times S^5$ manifold via its Penrose limit. The
pp-wave spacetime is a maximal supersymmetric manifold in which 
string theory, in the light-cone gauge, is exactly solvable 
\cite{1, 2}. 

On the other hand, we have the linear dilaton 
field as a profitable background for the non-critical string 
theory \cite{3}. This background has many prominent applications
in the string theory \cite{3, 4}. For example,  
it is a trusty method for reduction of the spacetime 
dimension without compactification. It also 
significantly modifies behaviors and evolutions of the 
strings and D-branes and their classical solutions \cite{5}.

In this paper we shall consider both of the above backgrounds 
simultaneously, i.e. we shall study an open string 
ending on a D$p$-brane
in the pp-wave spacetime with the linear dilaton field. 
Our calculations are in the light-cone gauge, and we shall investigate 
the solvability of the theory. For avoiding differential 
equations of order more than four we apply the dilaton field 
only in two dimensions of the brane. Therefore, the equations
completely are solvable. Due to the nature of the solutions of 
the worldsheet fields we have to do the quantization of the 
system by the symplectic method. 

This paper is organized as follows. In Sec. 2, our setup will be
fixed, and 
equations of motion and boundary conditions will be extracted. 
In Sec. 3, the solutions of the worldsheet fields will be given. 
In Sec. 4, the quantization of the system will be done. 
Sec. 5 is devoted to the conclusions and outlook.
\section{Setup and initial equations}

The pp-wave background includes a plane wave metric
with a constant R-R 5-form flux
\bea
&~& ds^2=-f^2
X^I X^I (dX^+)^2+2 d{X^+}dX^- + dX^I dX^I ,\;\;
I=1,2,\cdot\cdot\cdot,8
\nonumber\\
&~& F_{+1234}=F_{+5678}=2f.
\eea
Now we investigate behavior of an open string, attached to
a D$p$-brane, in the pp-wave spacetime which is smeared 
by a dilaton field $\Phi $. We use the light-cone 
coordinates $X^\pm=(X^9 \pm X^0)/\sqrt{2}$ 
with $X^+ = x^+ + 2\alpha' p^+ \tau$.
Therefore, the string action in these backgrounds is
given by 
\bea
S = & - &\frac{1}{4\pi\alpha'} \int_\Sigma d^2\sigma 
\sqrt{-h}\bigg{[}
h^{ab}\left( \partial_a X^I \partial_b X^I 
-f^2 X^I X^I \partial_a X^+ \partial_b X^+ 
+ 2\partial_a X^+ \partial_b X^- \right)
\nonumber\\
&+& \alpha'\Phi \;R^{(2)} \bigg{]},
\eea
where the string worldsheet $\Sigma$ has
the metric $h_{ab}$ with $h = \det h_{ab}$.
The scalar curvature of the worldsheet $R^{(2)}$ 
is constructed of the metric $h_{ab}$. 

Since an arbitrary dilaton field induces unsolvable 
equations, we consider a special dilaton which is 
a linear function of some coordinates along the brane.
That is, we apply the dilaton field $\Phi = a_k X^k$, 
where $a_k$ represents a constant vector field 
parallel to the $x^1x^2$-plane of the brane.
We supposed that dimension of the brane is $p \geq 2$
and its directions are 
$\{x^\alpha| \alpha =1,2, \cdot \cdot \cdot , p\}$. 
According to the diffeomorphism
symmetry of the action (2) we can choose a conformally
flat metric for the worldsheet, i.e. $h_{ab} (\sigma,\tau )=
e^{\rho(\sigma ,\tau)} \eta_{ab}$ with $\eta_{ab}={\rm diag} (-1,1)$. 
Presence of the dilaton field removes
the Weyl invariance, hence the scalar field $\rho (\sigma ,
\tau)$ is nonzero, and will be specified by the equations of 
motion. Adding all these together the string action,
ending on the D$p$-brane, takes the form 
\bea 
S =- \frac{1}{4\pi\alpha'} \int_\Sigma d^2\sigma 
\left(
\eta^{ab}\partial_a X^I \partial_b X^I 
+\mu^2 X^I X^I -4\alpha' p^+\partial_\tau X^-
+ \alpha' a_k X^k 
(\partial^2_\tau - \partial^2_\sigma)\rho\right),
\eea
where the mass parameter is $\mu = 2\alpha' p^+ f$, and for obtaining 
the last term we used the identity 
$\sqrt{-h}R^{(2)} = (\partial^2_\tau - \partial^2_\sigma )\rho$.

Vanishing of variation of the action exhibits equations
of motion for the fields $X^I$s and $\rho$,
\bea
&~& (\Box-\mu^2) X^{I'} = 0 ,\;\;\;\;\;\;\;\;\;\;\;\;
I' \in \{3,4,\cdot \cdot \cdot ,8\},\\
&~& (\Box-\mu^2) X^k
+ \frac {1}{2}\alpha'{a^k}\Box \rho\ =0,\;\;\;
k\in \{1,2\},\\
&~& a_k \Box X^k = 0,
\eea
where $\Box = -\partial^2_\tau + \partial^2_\sigma$.
Beside, variation of the action also yields the following 
boundary conditions for the open string
\bea
&~& (\partial_\sigma X^\alpha)|_{\sigma_0} = 0,\;\;\;
\alpha \in \{1,2,\cdot \cdot \cdot ,p\},\\
&~& \delta X^ i|_{\sigma_0} =0 ,\;\;\; 
i \in \{p+1,\cdot \cdot \cdot ,8\},\\
&~& ( \partial_\sigma \delta \rho)|_{\sigma_0} =0
\eea
where $\sigma_0 = 0,\;\pi $. These are conditions 
for the brane directions $X^\alpha$s, 
perpendicular directions to the brane $X^i$s except $X^9$,
and the scalar $\rho$, respectively.  

Now let consider a non-critical string theory, i.e. 
$a^2=a_k a^k \propto d-10 \neq 0$. Therefore, by combination   
of Eqs. (5) and (6) we acquire 
\bea
&~& (\Box-\mu^2) X^k + K^k_{\;\;l}X^l = 0,\\
&~& \Box \rho = \frac {2 \mu^2 }{\alpha' a^2}a_k X^k ,
\eea
where the matrix has definition 
$K_{kl}={\mu^2 {a_k a_l}}/{a^2}$. The Eq. (10)
elaborates that the worldsheet fields $X^k$s effectively
are stimulated by the potential 
\bea 
V(X)= -\frac{1}{2}K_{kl}X^kX^l 
+ \frac{1}{2}\mu^2 X^k X^k + V_0.
\eea 
In addition, Eq. (11) implies that, up to a constant factor, 
the linear dilaton is a source for the worldsheet 
scalar ``$\rho$''. According to the Eq. (3) this field interacts
kinematically with this source. 
We observe that simultaneous advent of the linear dilaton 
and the pp-wave backgrounds inspires the above potential
and source.

The Eq. (10) is decomposed to the following equations 
\bea
&~& \bigg{[} \Box - \left( \frac{\mu a_2}{a}\right)^2\bigg{]} X^1 
+\frac{\mu^2 a_1 a_2}{a^2} X^2=0 ,
\nonumber\\
&~& \bigg{[} \Box - \left( \frac{\mu a_1}{a}\right)^2\bigg{]} X^2 
+\frac{\mu^2 a_1 a_2}{a^2} X^1=0 .
\eea
By combining them we obtain a differential equation
and an algebraic one
\bea
&~& (\Box - \mu^2) X^1 = Y, \\
&~& X^2=-\frac {a^2}{\mu^2 a_1 a_2}Y - \frac{a_1}{a_2} X^1 ,
\eea
where the worldsheet variable $Y (\sigma , \tau )$
must satisfy the equations 
\bea 
&~& \Box Y (\sigma , \tau ) =0,\\
&~& (\partial_\sigma Y )|_{\sigma_0} =0 .
\eea
Thus, the variable $Y$ obeys the equations of the string coordinates     
along the brane worldvolume which lives in the flat spacetime.
According to Eq. (15) this variable (which is proportional to
the dilaton field) is not an independent worldsheet field, 
hence it cannot be interpreted as a new coordinate. 
\section{Solutions of the worldsheet fields}

Since the equations of 
$\{X^{I'}| I' = 3, 4, \cdot \cdot \cdot ,d-2\}$,
i.e. Eqs. (4), (7) and (8), 
are independent of the dilaton they have specific solutions, 
and hence for these equations we have 
\bea 
X^\alpha (\sigma , \tau ) &=& x^\alpha \cos \mu\tau
+ 2\alpha'p^\alpha \;\frac{\sin \mu\tau}{\mu}
\nonumber\\
&+& i\sqrt{2\alpha'}\sum_{n\neq 0}
\left( \frac{1}{\omega_n}\alpha^\alpha_n e^{-i\omega_n \tau}
\cos n\sigma \right),\;\;\;\alpha \in \{3,4,\cdot \cdot \cdot ,p\},
\nonumber\\
X^i (\sigma , \tau ) &=& \sqrt{2\alpha'}\sum_{n\neq 0}
\left( \frac{1}{\omega_n}\alpha^i_n e^{-i\omega_n \tau}
\sin n\sigma \right),\;\;\;i \in \{p+1,\cdot \cdot \cdot ,d-2\}.
\eea
The frequency $\omega_n$ is given by 
\bea
\omega_n = {\rm sign}(n)\sqrt{\mu^2+n^2}\;.
\eea
The second equation implies that the open string has 
been localized on the hyperplane $x^i =0$ without any 
center-of-mass momentum along these directions.

The general solution of the worldsheet variable $Y$ 
possesses the following expansion 
\bea 
Y (\sigma , \tau ) = y + y' \tau + i\sqrt{2\alpha'}
\sum^\infty_{n \neq 0} \left( \frac{1}{n}A_n 
e^{-in\tau}\cos n\sigma \right).
\eea 
The reality of this variable implies that  
$y$ and $y'$ also are real and $A^\dagger_n = A_{-n}$.
The solution of $X^1$, extracted from Eqs. (7)
and (14), finds the feature 
\bea 
X^1 = {\bar X}^1 - \frac{1}{\mu^2} Y,
\eea
where ${\bar X}^1$ is general solution of the homogeneous
equation, i.e. Eq. (14) without the right-hand side, with the 
boundary condition 
$(\partial_\sigma {\bar X}^1)|_{\sigma_0} =0$.
Therefore, we receive the solutions 
\bea 
X^1 (\sigma , \tau) &=& \frac{a^2_1 + a^2_2 \cos \mu\tau}{a^2}\;x^1
+ \frac{a_1 a_2}{a^2}(1-\cos \mu\tau) x^2
\nonumber\\
&+&  2\alpha' p^1 \;\frac{a^2_1 \tau + a^2_2 \;\frac{\sin \mu\tau}
{\mu}}{a^2}
+ 2\alpha' p^2 \;\frac{a_1 a_2}{a^2}\left( \tau - 
\frac{\sin \mu\tau}{\mu}\right)
\nonumber\\
&+&  i \frac{\sqrt{2\alpha'}}{a}
\sum^\infty_{n \neq 0}\left[\left( \frac{a_1}{n}\alpha^1_n e^{-in \tau}
- \frac{a_2}{\omega_n}\alpha^2_n e^{-i \omega_n \tau}\right)
\cos n\sigma \right],
\nonumber\\
X^2 (\sigma , \tau) &=& \frac{a_1 a_2}{a^2}(1-\cos \mu\tau) x^1
+ \frac{a^2_2 + a^2_1 \cos \mu\tau}{a^2}\;x^2
\nonumber\\
&+& 2\alpha' p^1 \;\frac{a_1 a_2}{a^2}\left( \tau - 
\frac{\sin \mu\tau}{\mu}\right)
+ 2\alpha' p^2 \;\frac{a^2_2 \tau + a^2_1 \;\frac{\sin \mu\tau}
{\mu}}{a^2}
\nonumber\\
&+&  i \frac{\sqrt{2\alpha'}}{a}
\sum^\infty_{n \neq 0}\left[\left( \frac{a_2}{n}\alpha^1_n e^{-in \tau}
+ \frac{a_1}{\omega_n}\alpha^2_n e^{-i \omega_n \tau}\right)
\cos n\sigma \right].
\eea
For obtaining these features for $X^1$ and $X^2$
we have used Eqs. (14) and (15) to replace the 
zero modes and oscillators of $Y$ as in the following
\bea 
&~& y=-\frac{\mu^2a_1}{a^2} (a_1 x^1 + a_2 x^2),
\nonumber\\
&~& y'=-\frac{2\alpha' \mu^2a_1}{a^2} (a_1 p^1 + a_2 p^2),
\nonumber\\
&~& A_n =-\frac{\mu^2a_1}{a}\alpha^1_n , \;\;\;
n \in {\mathbf Z}-\{0\}.
\eea 

For the case $\mu=0$, i.e. in the flat spacetime,
the zero-mode parts of the coordinates 
$X^1$ and $X^2$ change to the known results and,
by a redefinition of the oscillators, the oscillating 
parts of them also reduce to the expected forms.
For the next purposes we save the present  
oscillators. However, reality of these string coordinates 
inspires that $x^k$ and $p^k$ are real, and 
$(\alpha^k_{n})^\dagger = \alpha^k_{-n}$ for $k = 1,\; 2$.

For the worldsheet field $\rho (\sigma , \tau)$ we should  
solve Eq. (11) with the boundary condition (9),
which can be written as 
$(\partial_\tau \partial_\sigma \rho)|_{\sigma_0} =0$.
Therefore, the solution is given by
\bea 
\rho (\sigma , \tau) &=& \rho_0 + \rho' \tau + \rho'' \sigma
+ i \sqrt{2\alpha'} \sum^\infty_{n \neq 0}\left( \frac{1}{n}
\rho_n e^{-in\tau}\cos n\sigma\right) 
\nonumber\\
&+& \frac{\mu^2}{a^2}\bigg{[}\frac{1}{2\alpha'}
(a_1 x^1 + a_2 x^2)(-\tau^2 + \sigma^2)
- 2 (a_1 p^1 + a_2 p^2)\tau^2 \bigg{]}
\nonumber\\
&+& \frac{4 i}{a\sqrt{2\alpha'}}\bigg{[} 
\frac{a_2 (a_2 - a_1)}{a^2}\sum_{n \neq 0}
\left( \frac{1}{\omega_n}\alpha^2_n e^{-i\omega_n \tau}
\cos n\sigma\right)
\nonumber\\
&+& \mu^2 \sum_{n \neq 0}
\left( \frac{1}{n}f_n (\tau)\alpha^1_n e^{-in \tau}
\cos n\sigma\right)
\bigg{]},
\eea 
where the first line represents mode expansion of the 
general solution of the homogeneous equation 
$(\partial^2_\tau - \partial^2_\sigma)\rho =0$, 
while the other terms indicate 
the solution due to the right-hand side of Eq. (11).
The mode-dependent function $f_n (\tau)$ should 
satisfy the following equation 
\bea 
-\frac{d^2 f_n (\tau)}{d \tau^2} 
+2in \frac{d f_n (\tau)}{d \tau}=1 .
\eea 
It has the general solution 
\bea
f_n (\tau)=c_n - \frac{i}{2n}\tau + c'_n e^{2in\tau},
\eea 
where $c_n$ and $c'_n$ are integration constants.
Reality of the scalar field $\rho$ implies that $\rho_0$,
$\rho'$ and $\rho''$ are real and $\rho^\dagger_n =\rho_{-n}$, 
$c^\dagger_n =c_{-n}$, ${c'}^\dagger_n ={c'}_{-n}$.
Up to these conditions the coefficients 
$\{\rho_0 , \rho' , \rho'' , \rho_n\}$ 
remain arbitrary, and $c_n$ and ${c'}_n$ 
may be put away.

In fact, the dilaton also has a contribution 
to the energy-momentum tensor. The string action (3) is a 
conformal field theory which possesses the 
following energy-momentum tensor
\bea 
T_{\pm \pm} = \partial_\pm X^I \partial_\pm X^I
- \frac{1}{4}\mu^2 X^I X^I + 2\alpha' p^+ \partial_\pm X^-
-\alpha' a_k \partial^2_\pm X^k ,
\eea 
where $\partial_\pm =\frac{1}{2} (\partial_\tau \pm \partial_\sigma)$.
Vanishing of this tensor defines the equations 
of the coordinate $X^-$,
\bea 
&~& \partial_\tau X^- = - \frac{1}{4\alpha' p^+}\left(
\partial_\tau X^I \partial_\tau X^I + \partial_\sigma X^I
\partial_\sigma X^I -\mu^2 X^I X^I 
-\alpha' a_k (\partial^2_\tau + \partial^2_\sigma)X^k
\right),
\nonumber\\
&~& \partial_\sigma X^- = - \frac{1}{2\alpha' p^+}\left(
\partial_\tau X^I \partial_\sigma X^I 
-\alpha' a_k \partial_\tau \partial_\sigma X^k
\right).
\eea 
Therefore, the coordinate $X^-$ is expressed in
terms of the transverse coordinates $X^I$s.

Now we have exact solutions of all worldsheet fields 
$\{X^1 , X^2 , X^\alpha , X^i , X^+ , X^- ; \rho\}$.
Thus, one can easily calculate any quantity which is
a functional of them such as canonical momenta,
Hamiltonian and so on.
\section{Quantization of the system}

For the field $X^k$ the action (3) defines the conjugate momentum  
\bea 
P^k (\sigma , \tau )= \frac{1}{2\pi \alpha'}\partial_\tau 
X^k (\sigma , \tau ).
\eea
In terms of the modes we explicitly acquire  
\bea 
P^1 (\sigma , \tau ) &=& \frac{1}{2\pi \alpha' a^2}
\bigg{[} \mu a_2( -a_2 x^1 +a_1  x^2) \sin \mu\tau 
\nonumber\\
&+& 2\alpha' p^1 \left( a^2_1 + a^2_2 \cos \mu\tau \right)
+ 2\alpha' p^2 a_1a_2\left( 1- \cos \mu\tau\right)\bigg{]}
\nonumber\\
&+& \frac{1}{\pi a \sqrt{2\alpha'}} \sum^\infty_{n \neq 0}
\left[\left( -a_2 \alpha^2_n e^{-i\omega_n \tau}
+ a_1 \alpha^1_n e^{-in\tau}\right) \cos n\sigma \right],
\nonumber\\
P^2 (\sigma , \tau ) &=& \frac{1}{2\pi \alpha' a^2}
\bigg{[} \mu a_1( a_2 x^1 - a_1  x^2) \sin \mu\tau 
\nonumber\\
&+& 2\alpha' p^1 a_1 a_2\left( 1- \cos \mu\tau\right)
+2\alpha' p^2 \left( a^2_2 + a^2_1 \cos \mu\tau \right)
\bigg{]}
\nonumber\\
&+& \frac{1}{\pi a \sqrt{2\alpha'}} \sum^\infty_{n \neq 0}
\left[\left( a_1 \alpha^2_n e^{-i\omega_n \tau}
+ a_2 \alpha^1_n e^{-in\tau}\right) \cos n\sigma \right].
\eea
The components of the total momentum of the open string, i.e.   
$P^k_{\rm total}(\tau )=\int^\pi_0 d\sigma P^k (\sigma , \tau )$,
are given by 
\bea 
P^1_{\rm total} (\tau ) &=& \frac{1}{2 \alpha' a^2}
\bigg{[} \mu a_2( -a_2 x^1 +a_1  x^2) \sin \mu\tau 
\nonumber\\
&+& 2\alpha' p^1 \left( a^2_1 + a^2_2 \cos \mu\tau \right)
+ 2\alpha' p^2 a_1a_2\left( 1- \cos \mu\tau\right)\bigg{]}
\nonumber\\
P^2_{\rm total} (\tau ) &=& \frac{1}{2 \alpha' a^2}
\bigg{[} \mu a_1( a_2 x^1 - a_1  x^2) \sin \mu\tau 
\nonumber\\
&+& 2\alpha' p^1 a_1 a_2\left( 1- \cos \mu\tau\right)
+2\alpha' p^2 \left( a^2_2 + a^2_1 \cos \mu\tau \right)
\bigg{]}.
\eea
It is obvious that these are not constants of motion. 
This is purely an effect of the pp-wave spacetime,
i.e. $\mu \neq 0$, in which the linear dilaton field has 
generalized it. 

Since there are various frequencies in $X^k$ and $P^k$,
i.e. $\mu$, $n$ and $\omega_n$, the canonical quantization of the
system is very difficult. Thus, we use the symplectic form
\bea
\Omega = \int_0^\pi {\mathbf d}\sigma\;
\sum_{k=1}^2{\mathbf d}X^k\wedge {\mathbf d}P^k .
\eea 
This differential form can be verified by the constraint 
structures of the theories \cite{6}-\cite{8}.
Introducing the mode expansions of $X^k$ and $P^k$ exhibits 
this 2-form in terms of the modes  
\bea
\Omega = {\mathbf d}x^1 \wedge {\mathbf d}p^1 
+ {\mathbf d}x^2 \wedge {\mathbf d}p^2
- i \sum^\infty_{n =1}\left( \frac{1}{n} 
{\mathbf d}\alpha^1_{-n}\wedge {\mathbf d}\alpha^1_{n}
+\frac{1}{\omega_n} {\mathbf d}\alpha^2_{-n}\wedge 
{\mathbf d}\alpha^2_{n}\right).
\eea
As a check, for the situation of $\mu =0 $, 
this reduces to the symplectic form 
of open string in the flat spacetime.
We observe that the symplectic form is $\tau$-independent,
hence it defines some appropriate Poisson brackets. 
By changing these brackets via the rule 
$\{\;,\;\}_{\rm P.B.} \rightarrow -i[\;,\;]$
we receive the commutation relations 
\bea 
&~& [x^1 , p^1]=[x^2 , p^2]=i,
\nonumber\\
&~& [\alpha^1_m , \alpha^1_n]=m\delta_{m+n,0},
\nonumber\\
&~& [\alpha^2_m , \alpha^2_n]=\omega_m\delta_{m+n,0},
\nonumber\\
&~& [\alpha^1_m , \alpha^2_n]=0.
\eea 
All other brackets are zero. These commutators imply 
the following equal-time commutation relations  
\bea 
&~& [X^k (\sigma , \tau) \;,\; P^l (\sigma' , \tau)]=
\frac{i}{\pi}\delta^{kl}\sum^\infty_{n = -\infty}
\left( \cos n\sigma \cos n\sigma'\right),\;\;\;\;k,l \in \{1,2\},
\nonumber\\
&~& [X^k (\sigma , \tau) \;,\; X^l (\sigma' , \tau)]=0,
\nonumber\\
&~& [P^k (\sigma , \tau) \;,\; P^l (\sigma' , \tau)]=0.
\eea
The infinite series is a representation of 
$\delta (\sigma -\sigma')$ on the interval 
$(0 , \pi)$. The last two equations elaborate that 
the space part and momentum part of the 4-dimensional
sub-phase space are commutative.

As we mentioned for the flat spacetime (i.e. the case with $\mu=0$), 
by a redefinition of the oscillators,
the coordinates $X^1$ and $X^2$ and their conjugate 
momenta are modified to the known simple 
forms. In fact, for receiving simple and elegant features for 
the symplectic form (33) and the commutators (34) 
and (35), we did not recast this redefinition. 
Thus, Eqs. (33), (34) and (35) remain independent of the  
characteristic vector of the linear dilaton.
Another advantage of the present oscillators is that 
they are dimensionless.
\section{Conclusions and outlook}

In this paper an open string, which has been
attached to a D$p$-brane, in the pp-wave spacetime 
with the linear dilaton was
studied. For obtaining a solvable model we chose 
the characteristic vector of the dilaton only in
two directions of the brane. This form of the 
dilaton field also is suitable for the
light-cone gauge formalism. Presence of this field 
effectively deforms equations of the system 
and, due to the lack of the Weyl symmetry,
preserves the worldsheet scalar ``$\rho$'' as a dynamical 
field. Appearance of different frequencies in the 
solutions of the worldsheet fields and their conjugate 
momenta imposes the symplectic quantization on the system.
As expected, the directions of the brane which include
the dilaton field, similar to all other directions, remain 
commutative.

The equations of the string coordinates 
specify a differential equation of order four. 
Solutions of these equations and the corresponding 
boundary equations demonstrate that behavior of an 
open string is drastically influenced by the
combination of the dilaton and pp-wave backgrounds. 
We observed that this combination also is origin of a 
potential, and defines a source for the scalar 
``$\rho$'', too. However, because of the pp-wave background  
the string momentum is not constant of motion. The lack 
of this conservation is independent of the dilaton background.
Finally, besides the undetermined Fourier coefficients in the solution 
of the Weyl scalar ``$\rho$'', this field also possesses two 
mode-dependent constants which remain unknown. 

It is interesting to extend the setup with more components
of the characteristic vector of the dilation $a_k$,
then find out whether new results occur. In fact, by saving
$N$ components of this vector along the brane directions 
we receive some differential 
equations of the order $2N$. As another extension, the setup can be
extended to the superstring theory with two or more components
of the characteristic vector of the dilation.

\end{document}